\documentclass[conference]{IEEEtran}
\IEEEoverridecommandlockouts

\usepackage{geometry}
\def\BibTeX{{\rm B\kern-.05em{\sc i\kern-.025em b}\kern-.08em
    T\kern-.1667em\lower.7ex\hbox{E}\kern-.125emX}}

\usepackage{float}
\usepackage{xcolor}
\usepackage{epsfig, amsmath, amssymb}
\usepackage{wrapfig,lipsum,booktabs}

\usepackage{epstopdf}
\usepackage{balance}
\usepackage{color}
\usepackage{url}
\usepackage{pifont}
\usepackage{paralist}
\usepackage{enumitem}
\usepackage{balance}
\usepackage{comment}
\usepackage{multirow}
\usepackage{upgreek}
\usepackage[ruled,vlined]{algorithm2e}
\usepackage{cite}
\usepackage[fleqn]{nccmath}
\usepackage{epsfig,amsfonts,amsmath,amssymb, color,wrapfig}
\usepackage{epstopdf}
\usepackage{epsfig}

\usepackage[normalem]{ulem}

\usepackage{epsfig, subfigure, amsmath, amssymb}

\usepackage{epstopdf}
\usepackage{balance}
\usepackage{color}
\usepackage{url}
\usepackage{pifont}
\usepackage{paralist}
\usepackage{commath}
\usepackage{enumitem}
\usepackage{balance}
\usepackage{comment}
\usepackage{xcolor}
\usepackage{soul}
\usepackage{graphicx}
\usepackage{mwe}
\usepackage{hyperref}

\usepackage{pifont}
\usepackage{graphicx}
\usepackage{xcolor}
\usepackage{enumitem}
\usepackage{multicol}

\definecolor{red}{rgb}{1,0,0}
\definecolor{green}{rgb}{0,1,0}
\definecolor{blue}{rgb}{0,0,1}
\definecolor{aqua}{rgb}{0.7,0,0.7}
\definecolor{orange}{rgb}{1,0.64,0.06}

\def\BibTeX{{\rm B\kern-.05em{\sc i\kern-.025em b}\kern-.08em
    T\kern-.1667em\lower.7ex\hbox{E}\kern-.125emX}}

\begin{document}

\title{G-IDS: Generative Adversarial Networks Assisted Intrusion Detection System}

\author{
\IEEEauthorblockN{Md Hasan Shahriar\IEEEauthorrefmark{1}, Nur Imtiazul Haque\IEEEauthorrefmark{1}, Mohammad Ashiqur Rahman\IEEEauthorrefmark{1}, and Miguel Alonso Jr\IEEEauthorrefmark{2}}
\IEEEauthorblockA{\IEEEauthorrefmark{1}Analytics for Cyber Defense (ACyD) Lab, Florida International University, USA\\
	\IEEEauthorrefmark{2}School of Computing and Information Sciences, Florida International University, USA\\
	\{mshah068, nhaqu004, marahman, malonsoj\}@fiu.edu
	}
}

\maketitle

\begin{abstract}
The boundaries of cyber-physical systems (CPS) and the Internet of Things (IoT) are converging together day by day to introduce a common platform on hybrid systems. Moreover, the combination of artificial intelligence (AI) with CPS creates a new dimension of technological advancement. All these connectivity and dependability are creating massive space for the attackers to launch cyber attacks. To defend against these attacks, intrusion detection system (IDS) has been widely used. However, emerging CPS technologies suffer from imbalanced and missing sample data, which makes the training of IDS difficult. In this paper, we propose a generative adversarial network (GAN) based intrusion detection system (G-IDS), where GAN generates synthetic samples, and IDS gets trained on them along with the original ones. G-IDS also fixes the difficulties of imbalanced or missing data problems. We model a network security dataset for an emerging CPS technologies using NSL KDD-99 dataset and evaluate our proposed model's performance using different metrics. We find that our proposed G-IDS model performs much better in attack detection and model stabilization during the training process than a standalone IDS.
\end{abstract}

\section{Introduction}
\label{sec:Intro}
Cyber-physical systems (CPS) can be referred to as contemporary systems with assimilated computational and physical capabilities that can communicate with humans in new modalities~\cite{lee2008cyber}. These systems have an enormous impact on various domains like environmental monitoring, intelligent transportation, production system, smart grid, smart home, smart city, and the smart healthcare system. All these domains are dependent on the network as they require remote data transfer for sending data from sensors to actuators via a control center. Communication in the wide-open network makes the system vulnerable and creates a humongous attack space for adversaries~\cite{ suo2012security}.

The Internet of things (IoT), one of the essential sub-domain of CPS, has taken a significant technological advancement to a whole new level where data is the main force. IoT has opened up a new dimension in conjunction with actuators, electronics, sensors, software, and connectivity to enhance connection, collection, and data exchange. In spite of wide acceptability, almost 80\% IoT devices are vulnerable to a wide range of cyber attacks~\cite{smarthome2019}. They are susceptible to various kinds of attacks like man-in-the-middle, data and identity theft, distributed denial of service (DDoS), device hijacking  etc. To protect safety critical systems from the intruders, robust security measures must be taken into consideration to detect all kinds of known and unknown attacks~\cite{kolias2017ddos,rahman2019false, sikder2019aegis}. 
Intrusion Detection Systems (IDSs), responsible for inspecting network traffic and system data for malicious activities and issuing alerts, are the first and foremost part of the defense strategy in the CPS domain.  Having the proper knowledge of exact place and time where specific abnormalities are creating hazards in the system helps to mitigate the impacts by taking appropriate actions, and thus intrusion prevention systems come into the picture. The intrusion prevention system works simultaneously with an intrusion detection system to prevent the attacker from doing any harm to the system. 

Machine learning-based IDS can detect abnormalities in the system with substantial accuracy~\cite{aldweesh2020deep}. 
Though the emerging IT trends in CPS such as Industry 4.0, IoT, big data, and cloud computing are adding more momentum, they are introducing more vulnerabilities as well~\cite{kim2017review}. Besides, novel architectural compositions are adding complexity to the model due to unknown emergent behavior~\cite{torngren2018deal}. Individual IDS needs to be implemented to observe their interaction with this complex system, but insufficient data are limiting the model training. Moreover, most of these available datasets are imbalanced where different types of attack data are not available on a large scale compared to the normal data. Thus, to handle these limitations, we develop a generative adversarial network (GAN)-assisted IDS that can mitigate the dataset related limitations for all these emerging technologies. 

Our contribution to the paper can be summarized in the following points:
\begin{itemize}
\item We model an IDS using artificial neural network (ANN) that can be trained on any database with high accuracy.
\item We model an emerging CPS security dataset where standalone IDS is unable to predict with high accuracy.
\item Most importantly, we propose a new security framework, G-IDS, where GAN generates more training data to solve the imbalanced and missing data problems. Evaluating the performance of standalone IDS (S-IDS) and G-IDS, we find that G-IDS outperforms S-IDS in different ways.
\item We evaluate our proposed framework in a widespread network intrusion detection dataset KDD'99.
\end{itemize}

\noindent\textit{Organization:} The rest of the paper is organized as follows: We add sufficient background information, and our motivation towards this research focus in Section~\ref{sec:background_motivation}. The related works are discussed in Section~\ref{sec:related_work}. We introduce our proposed G-IDS framework in Section~\ref{sec:proposed_methodology}. Section~\ref{sec:technical_details} discusses the technical details of the framework, along with the complete analysis of our algorithm. In Section~\ref{sec:experiment}, we explain the evaluation setup and dataset along with the data pre-processing and data modeling steps.  The empirical analysis and finding are formulated in Section~\ref{sec:results}. At last, we conclude the paper in Section~\ref{sec:conclusion}. 
\section{Background and Motivation} \label{sec:background_motivation}
The term CPS refers to a new generation of systems with integrated computational and physical capabilities that can interact with human through state-of-the-art modalities. 
However, the seamless connections of IoT in CPS realm are opening the gates for the cyber attackers to launch different types of attack. To defend a CPS from being compromised, IDS is widely used. 
In this section we briefly discuss security aspects of CPS and the importance of IDS. We also give a short overview of GAN that we apply in our proposed solution. Finally, we discuss the motivation of this research.

\subsection{Security Challenges in CPS}

CPS is an amalgamation of computation, networking, and physical processes.  
IoT is another form of CPS which is a realm of connecting devices with sensing, computational, and actuating power. The emerging technologies, i.e., real-time data processing, machine learning, big data analytics, embedded systems, are shaping the IoT infrastructure nowadays. 

Moreover, each day, lots of sophisticated and novel applications are being implemented. According to the statistics, in~2020, there will be 20.4 billion connected IoT devices, and by~2026 it will be around three trillion\cite{iotdata}. The recent invention of 5G technology can take the internet speed to a whole new level. Although the emerging CPS possess a humongous amount of data, the security and privacy of these complex systems face a significant amount of challenges due to non-homogeneous nature with other known systems and lack of related data and experimentation \cite{choo2017emerging, newaz2020survey}.

CPS are adding new vulnerabilities due to: 1) replacement of traditional electro-mechanical relay-based controller by sophisticated software interaction-based microprocessor and operating systems, which enlarges the potential threat spaces, 2) growth of connected open-network consisting sensitive components, 3) public availability of internal system design and protocols, 4) radical increase in the number of functionalities, operations, and events, and 5) drastic surge of the technologically skilled cyber-criminals~\cite{humayed2017cyber,rahman2019novel,ahmed2018malicious}. 

\begin{figure}[t]
\centering
\includegraphics[width=0.90\columnwidth]{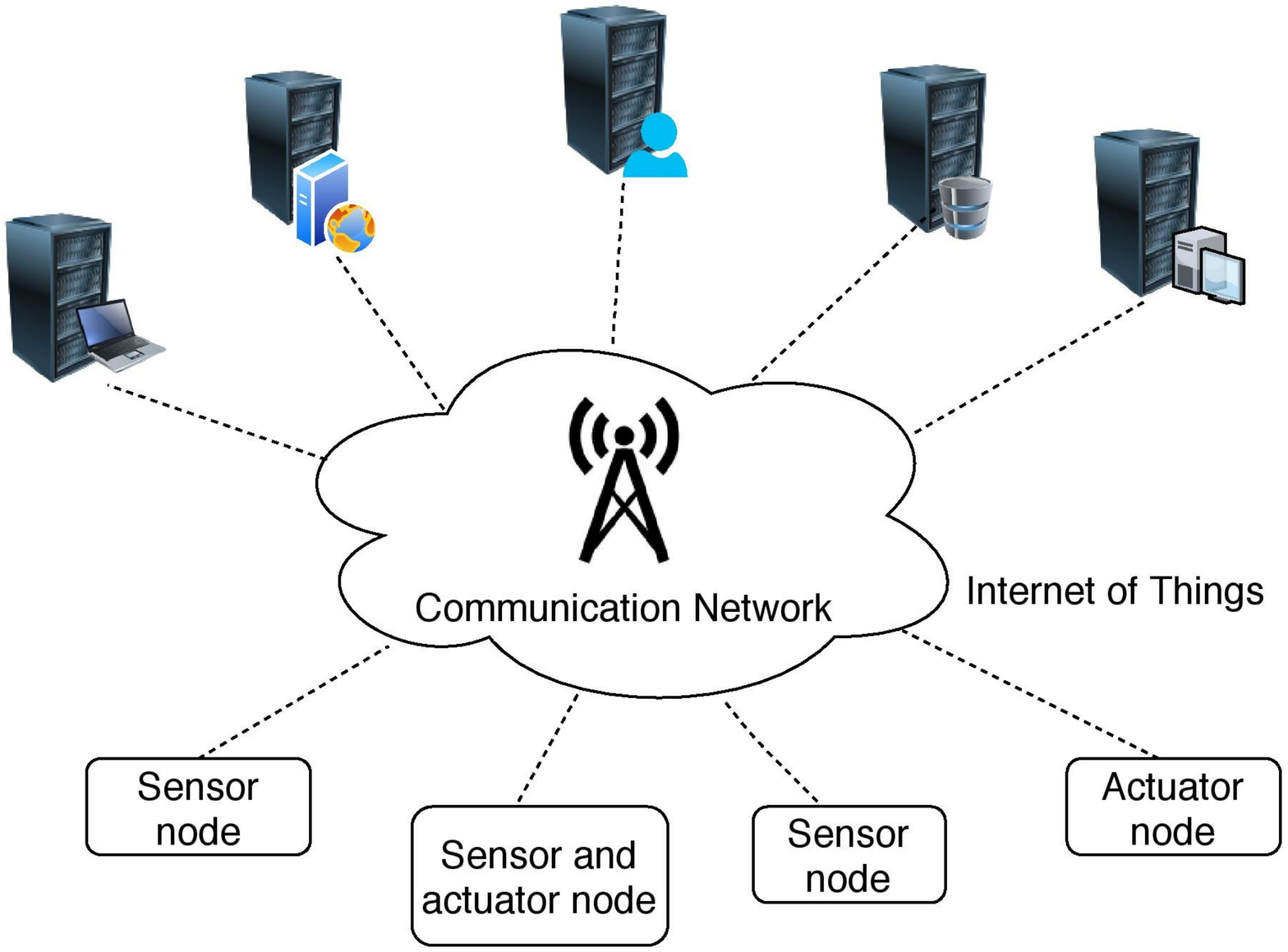} 
\vspace{-18pt}
\caption{Cyber-Physical System}
\vspace{-15pt}
\label{fig:cps}
\end{figure}

\subsection{Intrusion Detection Systems in CPS}
  
Along with the positive impacts of CPS in our lives, the risk of security breaches enhances parallelly. Moreover, IoT is being used in most of the critical infrastructures, which may cause a huge loss if an attacker gets control over them. An enormous amount of data is being processed and sent through the communication to keep the synchronization among all the equipment. A slight alteration in the data can cause severe destruction in the whole system. An IDS is a device or software application that monitors a network or system for harmful activities and rule violations. Thus, along with the device-level security, a dedicated IDS can ensure the high-level defense to the whole system. 

Different types of IDS are widely implemented in the CPS domain. Data mining, machine learning, rule-based model, statistical modeling, protocol model, signal processing model, etc. are used in CPS~\cite{anwar2017intrusion}. 
Among them, anomaly-based IDS is mostly used. Anomaly-based IDS learns the distribution of data in the training process. If the new coming data is not from the same distribution, it is detected as an anomaly. Whenever the system has new training data, it trains the IDS model and keeps updated with the novel attacks. Thus, the defense mechanism of anomaly-based IDS is a dynamic process and needs more data for each label to train its model correctly~\cite{abie2019cognitive}.

Other than the anomaly-based model, various types of intrusion detection techniques are also used in the defense mechanism. Different machine learning and deep learning-based algorithms are showing high detection accuracy~\cite{song2020novel, sikder20176thsense, makani2018taxonomy, neha2020sco}. 
For classification type IDS, supervised learning is widely used, and unsupervised learning is usually used for clustering based IDS~\cite{newaz2019healthguard}. Besides, different data mining techniques are also used to extract important features from the bulk amount of data~\cite{duque2015using}. Statistical analysis also shows an efficient performance in the anomaly detector from the historic data~\cite{amin2009rides}. Signal processing is also used to observe the pattern of the network and detect the abnormality \cite{vancea2015some}

\subsection{Generative Adversarial Networks (GAN)}

GAN was invented by  Goodfellow et. al  in~2014 \cite{goodfellow}. It is one of the most powerful and promising tool in deep learning. GAN estimates a generative model though adversarial approach. It consists of two independent models: generator $(G)$ and discriminator $(D)$. The generative model $G$ estimates the data distribution $p(g)$ over real data space $x$. Considering an input noise variable $p(z)$, the goal of $G$ is to generate new adversarial sample $G(z)$ that comes from the same distribution of $x$. One the other hand, the discriminator model $D$ returns the probability $D(x)$, that the given sample $x$ is from real data set rather generated by $G$. The ultimate goal of $G$ is to maximize the probability that $D$ would mistakenly predict generated data as real  one and for $D$ the goal is to do the opposite. Thus, $G$ and $D$ play a two player minmax game and at the end they reach to an unique solution.  The value function $V(G,D)$ is defined as follows:
$$
\min _{G} \max _{D} V(D, G)=$$
$$\mathbb{E}_{\boldsymbol{x} \sim p_{\text {data }}(\boldsymbol{x})}[\log D(\boldsymbol{x})]+\mathbb{E}_{\boldsymbol{z} \sim p_{\boldsymbol{z}}(\boldsymbol{z})}[\log (1-D(G(\boldsymbol{z})))]
$$

Thus, to achieve the optimum solution, these two game participants need to continually optimize themselves to improve their ability to reach the Nash equilibrium~\cite{nash1951non}.

\subsection{Research Motivation}

The emerging CPS domains are going to reshape the world through technological advancement. Due to the uniqueness of each CPS domain, the attack patterns and threats also differ from each other. Most importantly, an emerging CPS always lacks such types of database to train the IDS to detect the probable attacks. Even if there is a database, the number of samples of each type of attack may not be enough for the IDS to get trained perfectly and detect with higher accuracy. Moreover, IDS is facing novel zero-day attacks more often as the cyber attackers are much more sophisticated nowadays and launch new types of cyberattacks regularly.  

Thus, IDS can not be trained with high accuracy unless sufficient data is available. Besides, the size of normal data points is too large compared to the abnormal/attack data, which creates a data imbalance problem during the training.  Moreover, even if there is a significant amount of data available, the distribution of the data space may not be apparent to IDS due to some missing data. On the other hand,  GAN is a promising deep learning tool with the capability of learning the distribution of provided data and generates new similar samples. 

Thus, to solve all the aforementioned problems, a GAN assisted IDS framework can be used to generate new data for all the labels that need improvements in prediction. Moreover, a comprehensive framework is also needed to deal with the imbalanced and missing data-related issues of the emerging CPS technologies.
\section{Related Work}
\label{sec:related_work}

There is a recent trend in the scientific community to study GAN's potential application in the CPS domains. GAN has promising applications in the security perspective of a CPS. Whereas some authors implemented GAN to attack, some utilized it to make the system robust.

Chhetri et al. proposed a conditional GAN based model to observe important security requirements by analyzing the relations between the cyber and physical domains in a CPS~\cite{8715283}. Yin et al. proposed a framework based on GAN along with the Botnet detection model, which improves the performance of the detection mechanism of a most formidable attack retaining the primary characteristics of the original detection model~\cite{yin2018enhancing}. Seo et al. proposed a GAN-based model to reduce false positive rate in vehicle networks for enhancing driver's safety~\cite{seo2018gids}. {Some research works have attempted to find vulnerabilities of the systems using GAN.} Li et al. used LSTM-RNN in GAN to capture the distribution of the multivariate time series of the sensors and actuators to detect abnormal working conditions for a sophisticated six-stage Secure Water Treatment (SWaT) system~\cite{li2018anomaly}. Usama et al. demonstrated that ML models are vulnerable to adversarial perturbation in the network traffic~\cite{usama2019generative}. They mainly focused on deceiving machine learning-based IDS using the GAN-based adversarial model. Huang et al. proposed IDSGAN, which leverages a generator to transform original malicious traffic into adversarial malicious traffic examples~\cite{lin2018idsgan}.

{On the other hand, GAN based anomaly detection model is gaining popularity.} Ferdowsi et al. proposed a distributed adversarial network for providing a fully decentralized IDS for the IoT realm to detect anomalies, which is practical to conceal the user's sensitive data~\cite{ferdowsi2019generative}. Belenko et al. explored the security breaches of a large scale CPS using generative adversarial ANNs~\cite{belenko2018evaluation}. {GAN based models are also poplar to deceive detection models.} Hu et al. proposed a GAN based framework named MalGAN to generate adversarial malware samples that can bypass any black-box machine learning-based detection models~\cite{hu2017generating}.

{All of these works represents the application of GAN in CPS security.} However, none of these works deals with imbalanced and limited data for emerging CPS technologies. Unlike the works mentioned above, we propose a comprehensive framework to synthesize training data that can improve the performance of IDS in cyber attack detection. Even though we evaluated our framework on the network intrusion detection dataset with ANN-based IDS, it is compatible with any IDS and different sensors and network datasets.

\section{Proposed G-IDS Framework} 
\label{sec:proposed_methodology}

\begin{figure*}[!ht]
\centering
\includegraphics[scale=0.5]{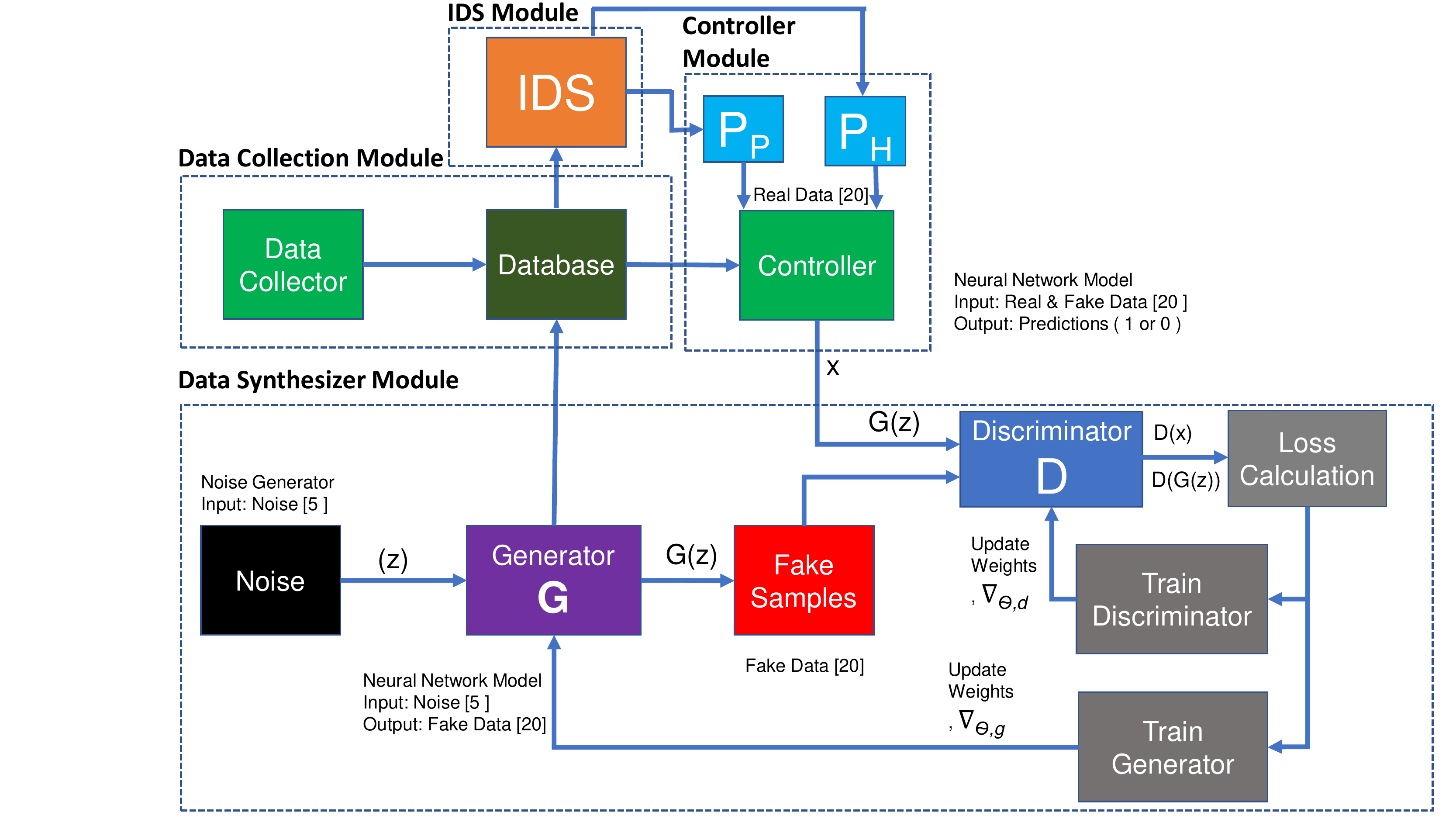}
\vspace{10pt}
\caption{Proposed framework of G-IDS}
\label{fig:ganids}
\end{figure*}

We divide our proposed G-IDS framework into four different segments:
1. Database module
2. Intrusion detection system module
3. Controller module, and
4. Synthesizer module.
Fig.~\ref{fig:ganids} shows the block diagram of our proposed framework. At first, the database module collects real-world intrusion detection data from its data collector. Database module may also get synthesized data from the generator of the data synthesizer module. All of these data are accumulated parallelly and continuously and stored in the database with different flags to distinguish data sources. The synthesized data 
are sub-categorized into \textit{pending} and \textit{synthetic} classes. \textit{Pending} data are uncommitted data that can be either accepted or deleted depending on the future decision of the controller module. On the other hand, data flagged as \textit{synthetic} are already verified and stay permanently in the database. We consider a hybrid database that contains only \textit{original} and \textit{synthetic} data samples.

An ML-based model in the intrusion detection systems module gets trained twice on the data available in the database. The first one is trained on the hybrid dataset only, while the other model gets trained on both hybrid and pending datasets. The controller evaluates the performances of these two models of the IDS. Based on the evaluation, the controller decides to reject/accept the pending samples in the database. The pending samples are accepted and appended to the database as \textit{synthetic} data if the model shows a better result after getting trained on them. Otherwise, the \textit{pending} samples are rejected and removed from the database. 

After that, controller module measures the detection rate for each class and makes a list of those for which the performance of the IDS falls below a certain threshold. Data samples for the weakly detected classes are sent to the data synthesizer module to generate more samples. Data Synthesizer module consists of a GAN, which generates new samples by learning the input data distribution. After generating potential samples, Data Synthesizer module feeds them in the database module with a \textit{pending} flag. The overall process is repeated until the label-wise performance of the IDS model becomes satisfactory to the controller. The Algorithm~\ref{gan_ids_algo} provides a comprehensive summary of the complete framework.  

\begin{algorithm*}[t]
\SetAlgoLined
initialization $Database$ = Data collected from DC\; 
 \For{true}{
    \begin{itemize}
        \item Train IDS model on $Hybrid Database$
        \item Update performance metrics, $PM_H$ 
        \item Controller lists all labels with $PM_H < PM_{TH}$ and sends them to the DSM
        \item Create $List\_ of\_weak\_labels$ for $PM_H < PM_{TH}$
   \end{itemize}
    \For{ each label  $l$ in  $List\_ of\_weak\_labels$}
    {
    \For{number of epoch}{

    \For{k steps}{
        \begin{itemize}
            \item Create a temporary copy of $Hybrid Database$ replacing label $l$ with 1 and the rest with 0. 
            \item Sample m noise { $z^{(1)},z^{(2)},..,z^{(m)} $} from noise distribution $p_g(z)$
            \item Sample batch of m examples { $x^{(1)},x^{(2)},..,x^{(m)} $} from $Hybrid Database$
            \item Update the discriminator model (D) by ascending its stochastic gradient:
            
            $$\Delta_{\theta_{d}} \frac{1}{m} \sum_{t=1}^{m} [log D(x^{(t)}) + log(1-D(G(z^{(t)})))]$$
        
        \end{itemize}
         }
         
        \begin{itemize}
            \item Sample batch of m noise { $z^{(1)},z^{(2)},...z^{(m)} $} from noise distribution $p_g(z)$
            \item Update the generator model by descending its stochastic gradient:
            
            $$\Delta_{\theta_{d}} \frac{1}{m} \sum_{t=1}^{m} log(1-D(G(z^{(t)})))]$$
        \end{itemize}
        }
        \begin{itemize}
            \item Generate $p$ samples for label $l$ and add them to $Database$ with flag \textit{p}
        \end{itemize}
        }
        \begin{itemize}
        \item Train IDS on $Database$ with pending samples
        \item Update performance metrics, $PM_P$
        \end{itemize}
        \eIf{$PM_P$ $>$ $PM_{H}$}{
        \begin{itemize}
            \item Accept pending (if any) samples for each label
            \item Updates flags from \textit{pending} to \textit{synthetic}
            \item add them to $Hybrid Database$
        \end{itemize}}{
    \begin{itemize}
        \item Remove pending samples from $Database$
    \end{itemize}
        } 
}
 \caption{Proposed Algorithm of G-IDS}
 \label{gan_ids_algo}
\end{algorithm*}

\section{G-IDS Technical Details} 
\label{sec:technical_details}

This section provides the technical description of the four modules of the G-IDS framework. 

\subsection{Database Module (DM)}
As shown in Fig.~\ref{fig:ganids}, DM consists of two important elements: data collector and database. 

\subsubsection{Data Collector (DC)}
In the proposed framework, we consider that the IDS is a network intrusion detector. Thus, DC collects real-world network data by capturing network packets. The packets are processed before feeding into the database. A packet capture application (i.e., Wireshark) runs within DC to collect packets and extract features from them. The responsibilities of DC is categorized into two types:     
\begin{itemize}
\item Data Collection:
     Data collection is the initial stage of the network-based IDS, as shown in Fig.~\ref{fig:ganids}. The communication between the sensors and the actuators of a CPS generates a vast amount of network data, which mostly contains normal/good data. However, an adversary can try to compromise the sensors or the communication channels to launch a cyber attack. Thus, to make the system robust against any kind of cyber attack, DC works along with a firewall to label those packets. After collection, DC does some pre-determined prepossessing on them to make them feedable to ML models.

\item Data Preprocessing: 
  As the network packet analyzer collects a massive chunk of data from the network, some prepossessing is mandatory to gather information from those packets. In our proposed model we consider the following tasks:\par
        \textbf{Encoding:} Label encoding is the technique of encoding categorical values in the dataset. It transforms each category of a particular feature with a value between 0 to $n$-1, where $n$ is the number of distinct categories of that feature. 
        
        \textbf{Feature Scaling:} Feature scaling is a method used to normalize the range of features.
         A dataset containing features that are highly varying in magnitudes, units, and range can become a critical problem for various ML algorithms that use Euclidean distance between two data points in their computations. To overcome this problem, we apply standard scaling to normalize the features. 
        
        \textbf{Feature Extraction:} Feature extraction (also known as feature reduction) remodels the high-dimensional space to fewer dimensions, where the transformation could be linear or non-linear. It helps to eliminate the redundant variables and makes the model a simpler one. 
        
\end{itemize}

\subsubsection{Database (DB)}
 Database stores the collected data from DC and DSM, along with the labels and corresponding flags. Each label represents the type of data (i.e., normal/attack). In the case of attack data, the label shows the type of attack. The flag bit for each sample indicates the status for that particular data as explained below:
\begin{itemize}
    \item \textit{\textbf{Original:}} Data collected from DC are flagged as \textit{original}. 
    \item \textit{\textbf{Pending:}} As mentioned in Section~\ref{sec:proposed_methodology}, DSM generates more data to improve the performance of IDS. However, due to the uncertainty of the GAN, every batch of the generated data does not guarantee the improvement of the performance of IDS. Hence, the generated data are flagged as \textit{pending} in the first place, and the controller does the further inspection. 
    \item \textit{\textbf{Synthetic:}} If the \textit{pending} data contribute to the improvement in the performance of IDS, the controller changes their flags from \textit{pending} to \textit{synthetic} and stores them permanently in the database. Whenever, the \textit{pending} data convert to \textit{synthetic}, they become the part of the hybrid dataset.
\end{itemize}

Table~\ref{tab:example_database} demonstrates an example database for some samples along with the labels and flags before encoding. 

\subsection{Intrusion Detection System Module (IDSM)} The core part of IDSM is an ML-based IDS which, in this case, is a multi-layer ANN-based model. It consists of 4 layers, where the input and output layers have 20 and 10 nodes, respectively, and both of the hidden layers consist of 50 nodes. 
IDS is trained 
twice to calculate the following two performance metrics which are useful to evaluate the synthesized data-

\begin{table}[t]
\vspace{-6pt}
\caption{An example database}
\label{tab:example_database}
\vspace{-3pt}
\centering
\begin{tabular}{|l|l|l|l|l|l|}
\hline
\textbf{Duration}       & \multicolumn{1}{l|}{\textbf{Protocol Type}} & \multicolumn{1}{l|}{\textbf{Service}} & \multicolumn{1}{l|}{\textbf{\dots}} & \multicolumn{1}{l|}{\textbf{Label}} & \textbf{Flag} \\ \hline
0                       & tcp                                         & http                                  & \dots                               & normal                              & original      \\ \hline
0                       & icmp                                        & ecr\_i                                & \dots                               & smurf                               & original      \\ \hline
18848                       & tcp                                        & telnet                                & \dots                               & normal                               & pending      \\ \hline
199                       & tcp                                        & telnet                                & \dots                               & spy                               & synthetic      \\ \hline

\multicolumn{1}{|c|}{\vdots} &\vdots                                           &\vdots                                     & \vdots                               & \vdots                                   & \vdots       \\ \hline
18848                   & tcp                                         & telnet                                & …                               & normal                              & synthetic     \\ \hline
\end{tabular}
\vspace{-12pt}
\end{table}

\textbf{Performance metric without pending data ($\mathit{PM_H}$):}
After collecting data from the data collection module, IDS trains the model on the hybrid database containing only {original} and {synthetic} data. Then, IDS evaluates the label-wise performance metrics and stores in $\mathit{PM_H}$, which is the current best possible performance of the module.

\textbf{Performance metric with pending data ($\mathit{PM_P}$):}
After obtaining synthesized data with the \textit{pending} flag, IDS re-train the model on the hybrid database, including the pending data, to evaluate the additional contribution of the pending data points. The performance metric is stored in $\mathit{PM_P}$. Both of these evaluation metrics are available to the controller module based on which it takes a further decision on pending data. 
\subsection{Controller Module (CM)}
CM performs two important tasks in parallel.  

\textbf{Sending data synthesize request to the DSM:} Firstly, the controller analyzes $\mathit{PM_H}$ and compare it with the $\mathit{PM_{TH}}$, which is the minimum threshold for the performance metric. Any class with an individual performance score below the $\mathit{PM_{TH}}$ is considered as the weak class, which needs the aid of DSM in the generation of new samples to improve the detection rate. Each time the CM sends a request for one label with the hybrid database and continues for all the remaining weak ones.

\textbf{Evaluating pending data requests from DSM:} CM analyzes the evaluation metric $\mathit{PM_P}$ and $\mathit{PM_H}$ to update the database. Comparing these two metrics, CM suggests the database to remove the \textit{pending} samples for a specific label if the performance degrades with them. On the other hand, if the performance improves, CM recommends accepting the pending data by updating the \textit{pending} flags to \textit{synthetic}.

\subsection{Data Synthesizer Module (DSM)}
The core part of the DSM is a GAN-based model. Fig.~\ref{fig:ganids} shows the DSM, which consists of two ML models: 1. generator (G) and 2. discriminator (D). The controller selects one class and sends the whole database to the DSM, mentioning $1$ as the label of that specific class and $0$ for the rest of the classes. Thus, for DSM, it becomes a binary classification model and generates more samples after successful training. We model both the generator and discriminator using four-layer ANN. Each of the hidden layers has 50 nodes. The input and output layers of discriminator have 20 and 1 nodes, respectively. Besides, the generator takes a noise (\textit{latent space}) as an input and produces artificial samples. Thus the input and output layers of the generator have 5 and 20 nodes, respectively. Once the GAN training is complete, it generates new samples with higher accuracy and asks the DB to add them with \textit{pending} flags. 

\section{Evaluation Methodology} \label{sec:experiment}
In this section, we introduce the experimental setup and necessary evaluation metrics to assess our proposed G-IDS framework's performance. 

\subsection{Dataset}
The intrusion detection dataset of emerging CPS carries a lot of benign data comparing to attack data. Thus, we resemble the empirical data-limitation scenario by modifying KDD'99 dataset~\cite{tavallaee2009detailed} and use it to evaluate our model by analyzing different attack detection.  

The dataset contains ten different labels, which can be categorized into a. Denial of Service Attack (DoS), b. User to Root(UTR) Attack, c. Remote to Local Attack (R2L), and d. Probing Attack (PA), etc~\cite{ozgur2016review}. Table~\ref{tab:ganidstable} demonstrates the number of supporting data for all attack labels. The characteristics of our dataset can be formalized as follows:

\begin{table}[]
\centering
\caption{Class labels and the number of samples appearing in the dataset}
\label{tab:my-table}
\resizebox{0.45\textwidth}{!}{%
\begin{tabular}{|l|l|l|l|l|}
\hline
\textbf{Label} &\textbf{Attack} & \textbf{Original Data} & \textbf{Class} \\ \hline
1  & Back  & 2003 &  DoS      \\ \hline
2  & Ipsweep probe  & 1047 & PA      \\ \hline
3 & Perl u2r & 1070 & UTR  \\ \hline
4 & Normal & 9707 & Normal  \\ \hline
5 & Neptune dos & 1070 & DoS  \\ \hline
6 & Smurf dos & 840 & Do
S  \\ \hline
7 & Spy r2l & 1389 & R2L  \\ \hline
8 & Warezclient r2l & 779 & R2L  \\ \hline
9 & Warezmaster r2l & 820 & R2L  \\ \hline
10 & Phf & 550 & R2L  \\ \hline
\end{tabular}%
}
\vspace{-15pt}
\end{table}

\subsubsection{Multiclass}

Our dataset in consideration is a multiclass dataset that contains nine different types of attacks from four different classes and one normal label, as shown in Table~\ref{tab:my-table}. 

\subsubsection{Sparse}
Sparsity is a process of determining the number of missing values present in the dataset. A variable with sparse data is one in which a relatively high percentage of the variable cells do not contain actual data. Data sparsity is widely considered as a key cause of unsatisfactory classification accuracy. Our dataset is purely cleaned one with no missing attributes and records. However, sparsity calculation using the following equation returns 70\% sparsity, which is basically produced due to one-hot encoded features. Data sparsity can be calculated using:
$$Sparsity= 1-\frac{non\_zero\_count(X)}{number\_of\_elements(X)}$$
where X denotes the dataset.

\subsubsection{Imbalanced}
Imbalanced data hinders the performance of a classification problem. The number of observations per class is not equally distributed, and there is an enormous amount of observations for a particular class (majority class) and significantly less number of samples for one or more other classes (minority classes). Our dataset is an example of an imbalanced dataset as the number of benign examples dominate the number of attack examples. 

\subsection{Data Preprocessing}

Before training the model, some pre-processing is required to be performed. After a complete analysis, we perform the following techniques for data pre-processing.

\subsubsection{Encoding}

To deal with three categorical input features (namely, protocol\_type, service, and flag) and one output feature, we encode the corresponding data by applying the label encoding technique.

\subsubsection{Feature Scaling}
Feature scaling is an essential step to deal with local optima, and skewness towards particular features. It also facilitates the ML-based IDS with faster training. We apply standard scaling, which replaces the values by their Z-scores. Z-score of a data point represents the measure of deviation from the mean value. z-score, $z_i$, is calculated using the following equation:
$$z_i=\frac{x_i-x_{mean}}{x_{std}}$$
where $x_i$ represents the value of each feature,  $x_{mean}$ is the average value of the feature, and  $x_{std}$ denotes the standard deviation.

\subsubsection{Feature Extraction}
To reduce the number of features, we utilize principal component analysis (PCA)-based feature extraction method. PCA is an unsupervised, non-parametric statistical procedure that computes a new set of variables ("principal components") and expresses the data in terms of these new variables and uses an orthogonal transformation to transform a set of observations of possibly correlated variables~\cite{wold1987principal}. Mathematically, PCA is calculated via linear algebra functions called eigendecomposition, a process of factorization of a matrix into a canonical form.

For selecting the number of principal components, we use the explained variance ratio, representing the amount of information retained after applying PCA. The total variance is the sum of variances of all individual principal components. We use PCA on all the original 42 features and shrink the dimension into 20 elements, keeping the total explained variance  91.14\%. 

\subsection{Data Modeling}
To mimic an emerging CPS, we have taken only 5000 samples for training, and the rest of the data are used for testing purposes. As the dataset in consideration is imbalanced, the training data set loses some labels if the dataset is randomly split. To overcome this problem, we randomly select data from all available classes. 
\subsection{Evaluation}
We have considered the following metrics for evaluating the performance of our model.
\subsubsection{Precision}
Precision is the ratio of correctly predicted positive observations to the total positive observations. 

\subsubsection{Recall} Recall is the ratio of correctly predicted positive observations to the total actual positive observations. 

\subsubsection{F1 score} F1 score is a weighted average of precision and recall that takes both false positives and false negatives into account. In the case of unequal distribution of false Negative and false Positive values, F1 score can infer better comprehension of the model's performance. 

\subsubsection{Confusion Matrix}
Confusion matrix is a table that intends to present the count of correct and incorrect predictions made by each class. It gives a clear insight into the prediction and shows the classification/misclassification type (e.g., True Positive, True Negative, False Positive, False Negative). Thus, the performance of an IDS can be analyzed using these four important key factors as shown in Table~\ref{tab:metric}.

\begin{figure*}[t]
    \begin{center}
        \subfigure[]
        {
        \label{eval_5}
            \includegraphics[scale=0.48, keepaspectratio=true]{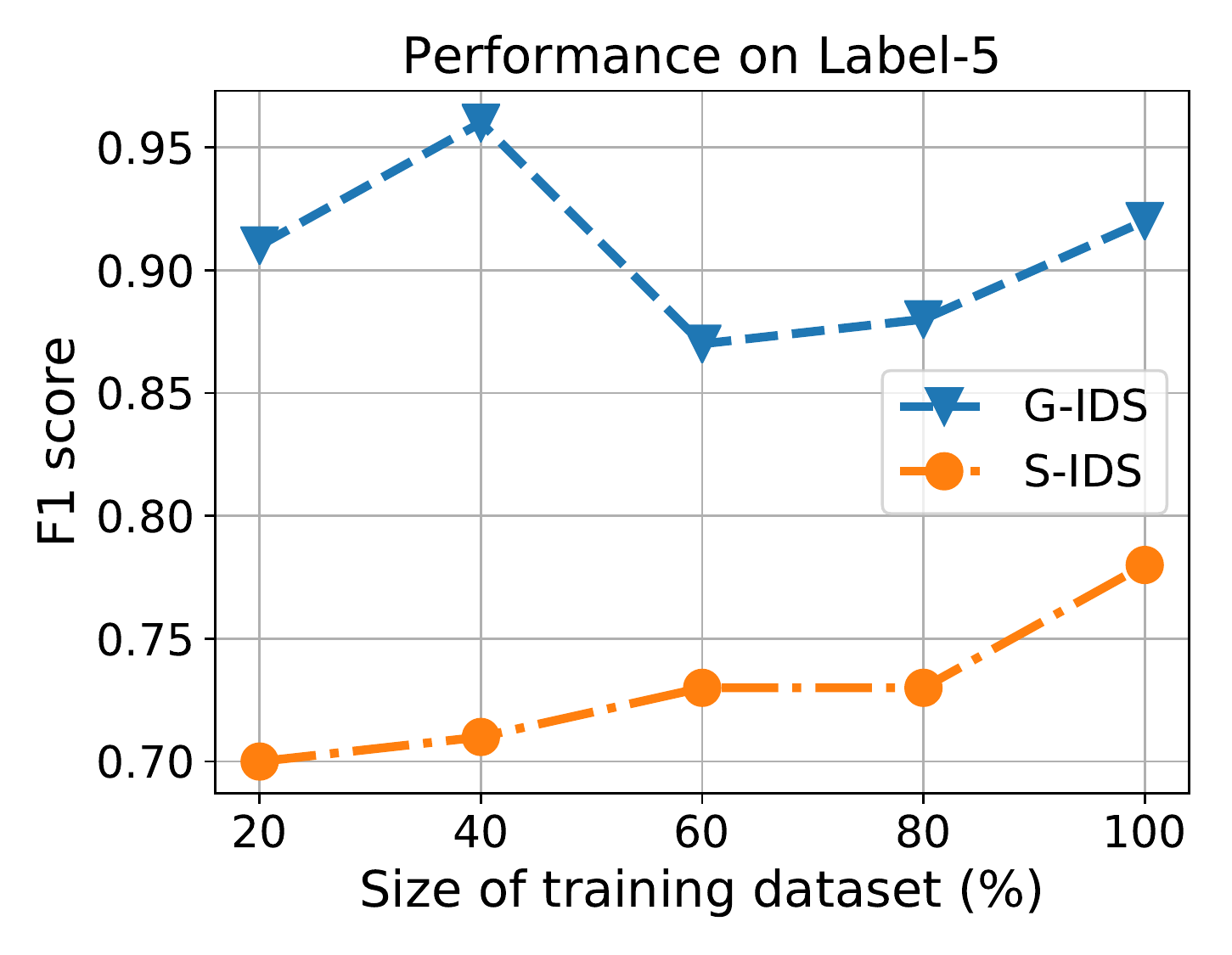}
        }
        \subfigure[]
        {
        \label{eval_6}
            \includegraphics[scale=0.48, keepaspectratio=true]{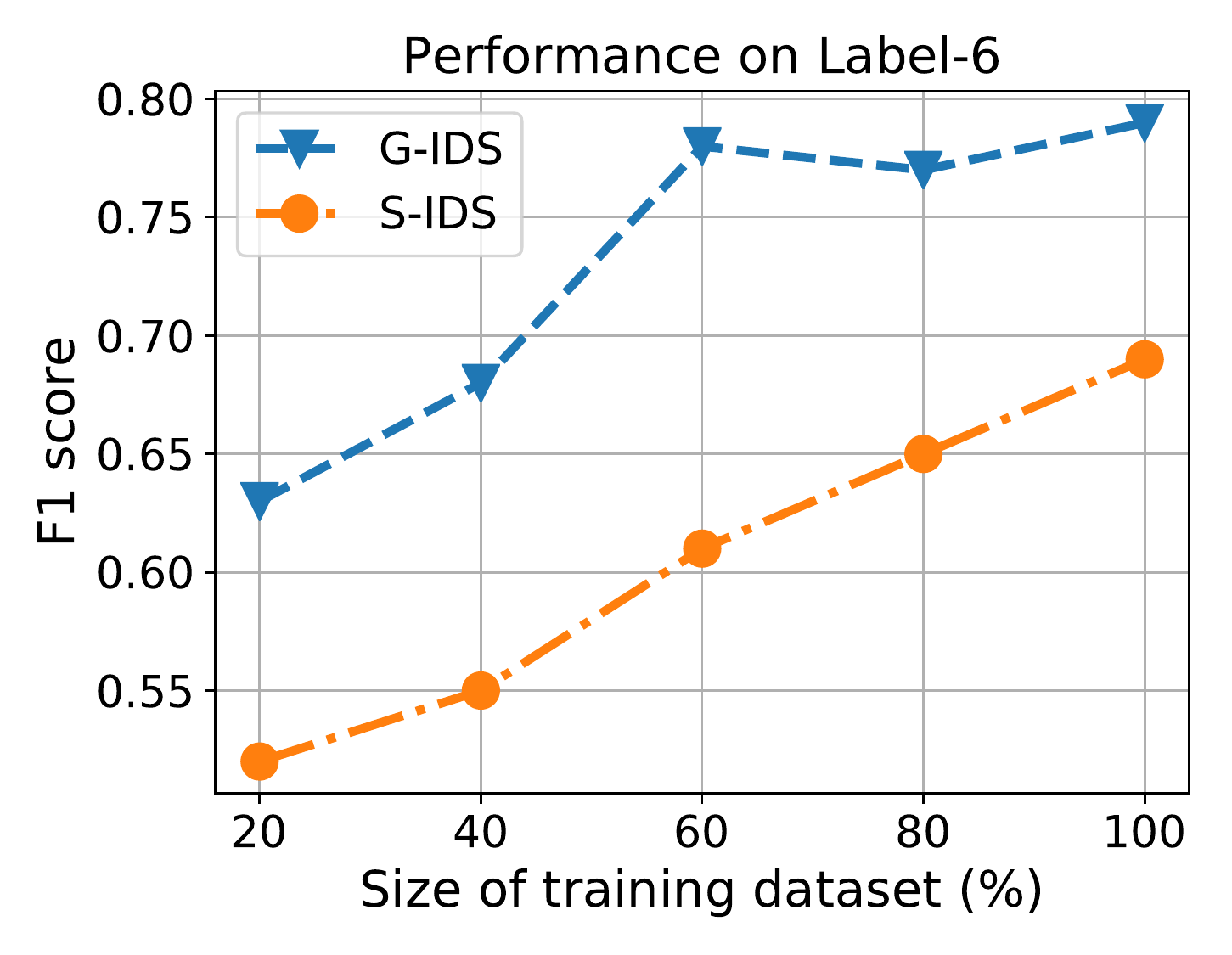}
        }
        \subfigure[]
        {
        \label{eval_9}
            \includegraphics[scale=0.48, keepaspectratio=true]{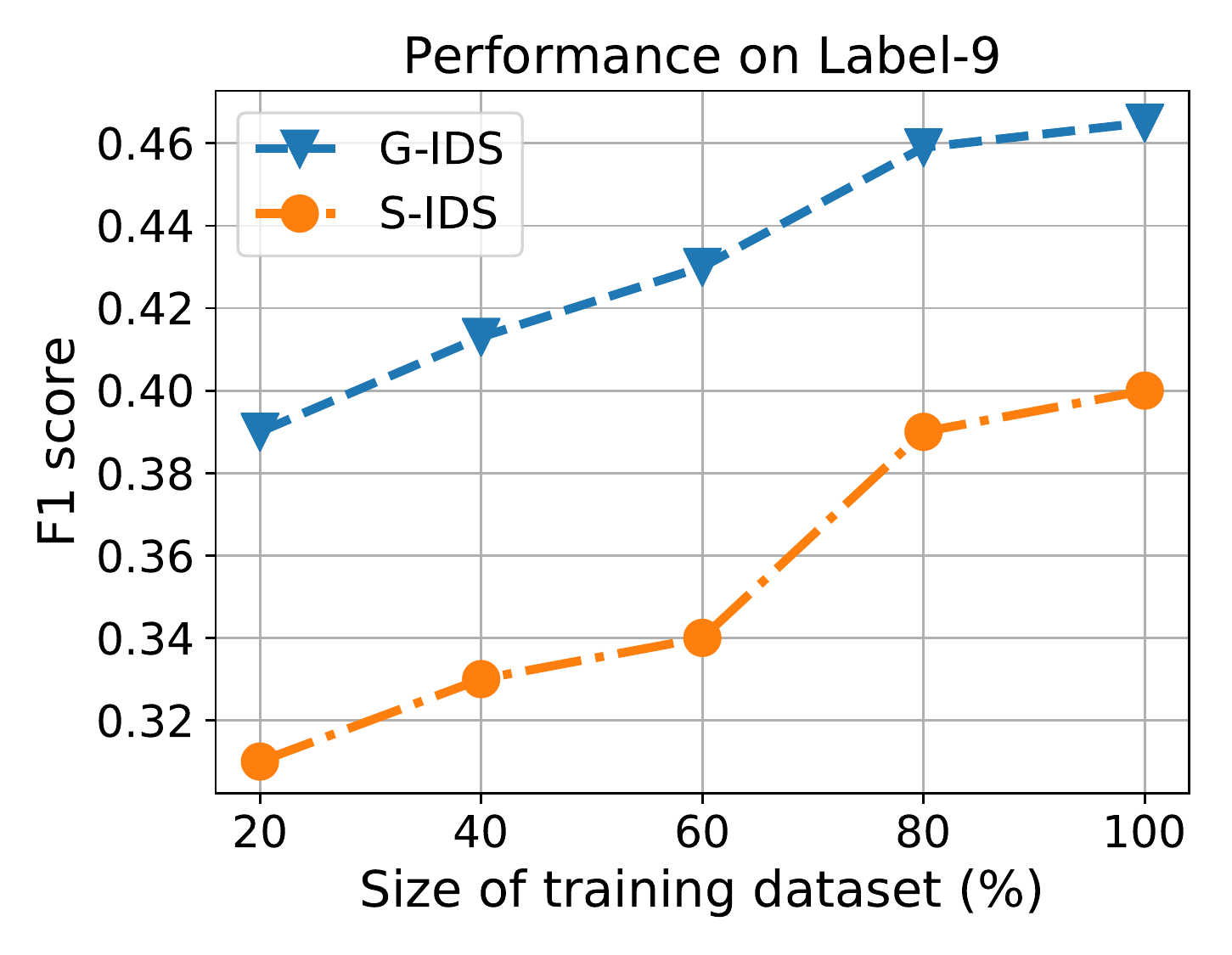}
        }
        \subfigure[]
        {
        \label{fig:overl_all}
            \includegraphics[scale= 0.48]{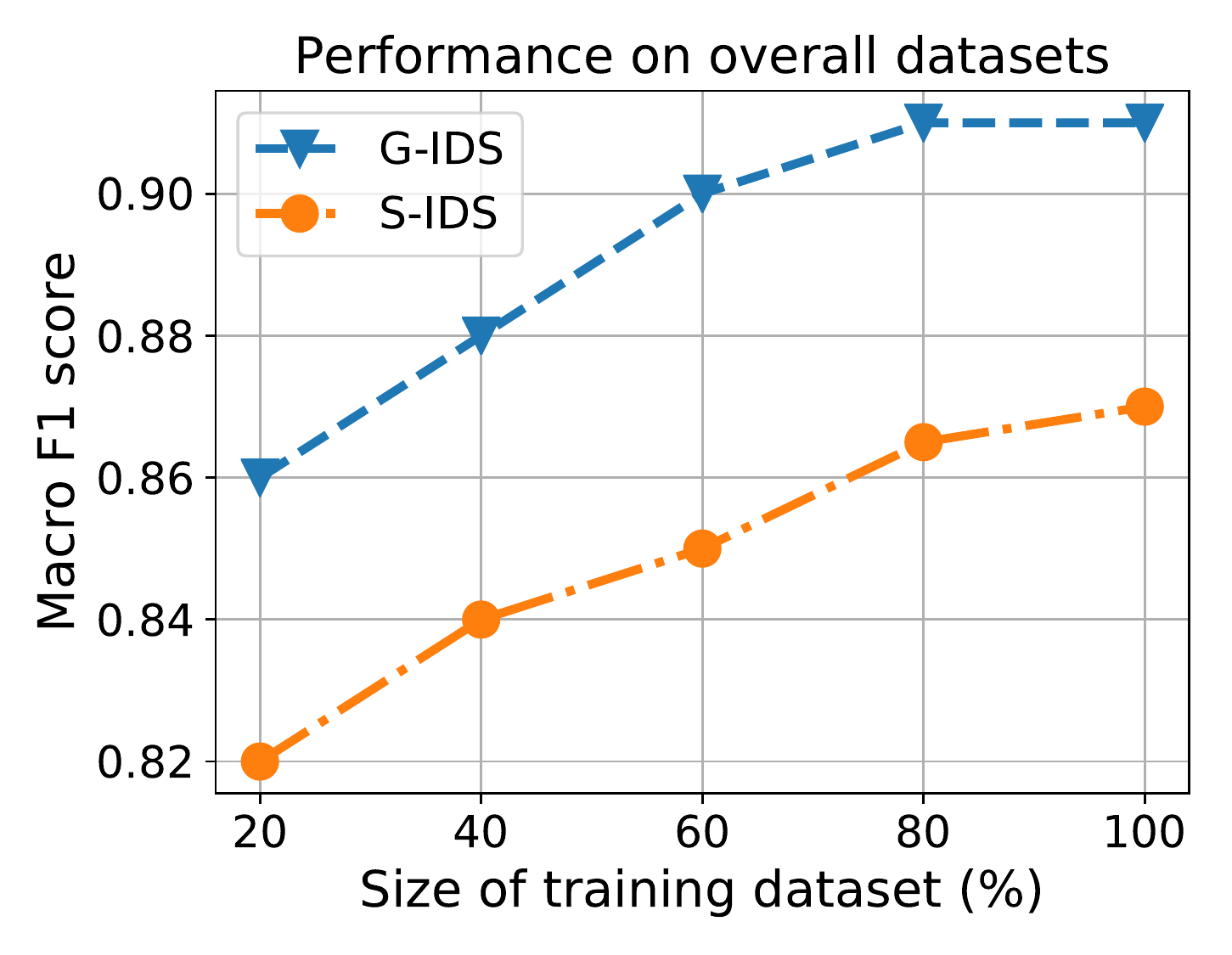}
        }
    \end{center}
    \vspace{-10pt}
    \caption{Performance evaluation of G-IDS and S-IDS in the prediction (F1 score) with respect to the size of training dataset for (a) label 5, (b) label 6, (c) label 9, and (d) overall dataset.}
    \label{fig:eval_labels}
\end{figure*}

\begin{table*}[t]
\caption{Performance metrics of S-IDS and G-IDS when trained with 40\% of original dataset}
\centering
\label{tab:ganidstable}
\resizebox{0.65\textwidth}{!}{%
\begin{tabular}{|c|c|c|c|c|c|c|c|}
\hline
\multirow{2}{*}{\textbf{Labels}} & \multicolumn{3}{c|}{\textbf{S-IDS}} & \multicolumn{3}{c|}{\textbf{G-IDS}} & \multirow{2}{*}{\textbf{Support}} \\ \cline{2-7}
 & \textbf{Precision} & \textbf{Recall} & \textbf{F1} & \textbf{Precision} & \textbf{Recall} & \textbf{F1} &  \\ \hline
1 & 0.99 & 0.99 & 0.99 & 0.98 & 0.99 & 0.99 & 2003 \\ \hline
\textit{\textbf{2}} & \textit{\textbf{0.83}} & \textit{\textbf{0.90}} & \textit{\textbf{0.85}} & \textit{\textbf{0.90}} & \textit{\textbf{0.95}} & \textit{\textbf{0.93}} & \textit{\textbf{1047}} \\ \hline
3 & 1.00 & 1.00 & 1.00 & 1.00 & 1.00 & 1.00 & 1070 \\ \hline
4 & 1.00 & 0.98 & 0.99 & 1.00 & 0.99 & 0.99 & 9707 \\ \hline
\textit{\textbf{5}} & \textit{\textbf{0.55}} & \textit{\textbf{0.98}} & \textit{\textbf{0.71}} & \textit{\textbf{0.93}} & \textit{\textbf{0.99}} & \textit{\textbf{0.96}} & \textit{\textbf{840}} \\ \hline
\textit{\textbf{6}} & \textit{\textbf{0.38}} & \textit{\textbf{0.98}} & \textit{\textbf{0.55}} & \textit{\textbf{0.52}} & \textit{\textbf{0.97}} & \textit{\textbf{0.68}} & \textit{\textbf{1389}} \\ \hline
7 & 1.00 & 1.00 & 1.00 & 1.00 & 1.00 & 1.00 & 2805 \\ \hline
8 & 1.00 & 1.00 & 1.00 & 1.00 & 1.00 & 1.00 & 779 \\ \hline
\textit{\textbf{9}} & \textit{\textbf{0.20}} & \textit{\textbf{0.98}} & \textit{\textbf{0.33}} & \textit{\textbf{0.27}} & \textit{\textbf{0.90}} & \textit{\textbf{0.41}} & \textit{\textbf{820}} \\ \hline
10 & 0.98 & 0.99 & 0.99 & 0.98 & 0.99 & 0.99 & 550 \\ \hline
\end{tabular}%
}
\end{table*}

\begin{figure*}[!ht]
    \begin{center}
        \subfigure[]
        {
            \includegraphics[scale=0.45, keepaspectratio=true]{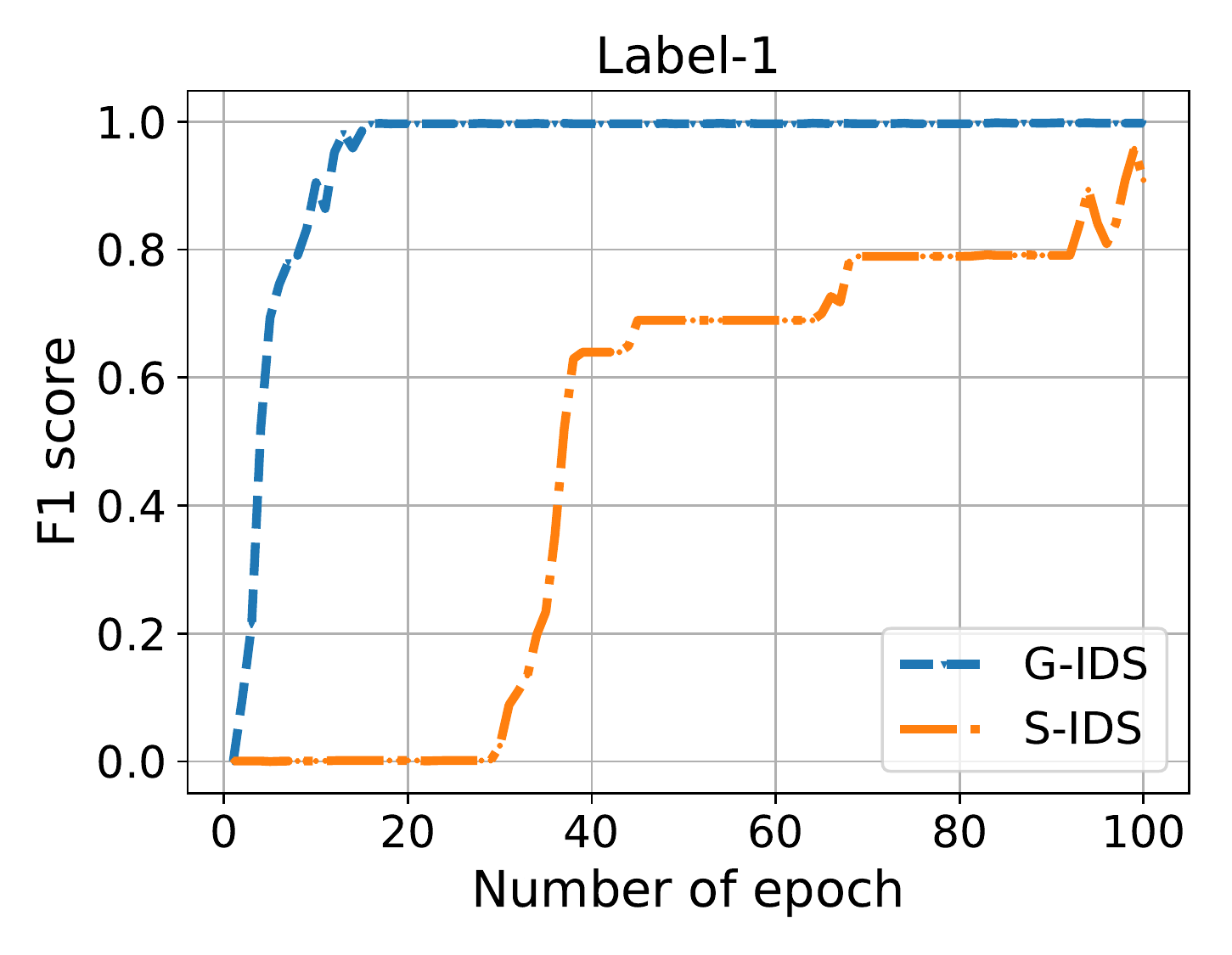}
        }
        \vspace{-5pt}
        \subfigure[]
        {
            \includegraphics[scale=0.45, keepaspectratio=true]{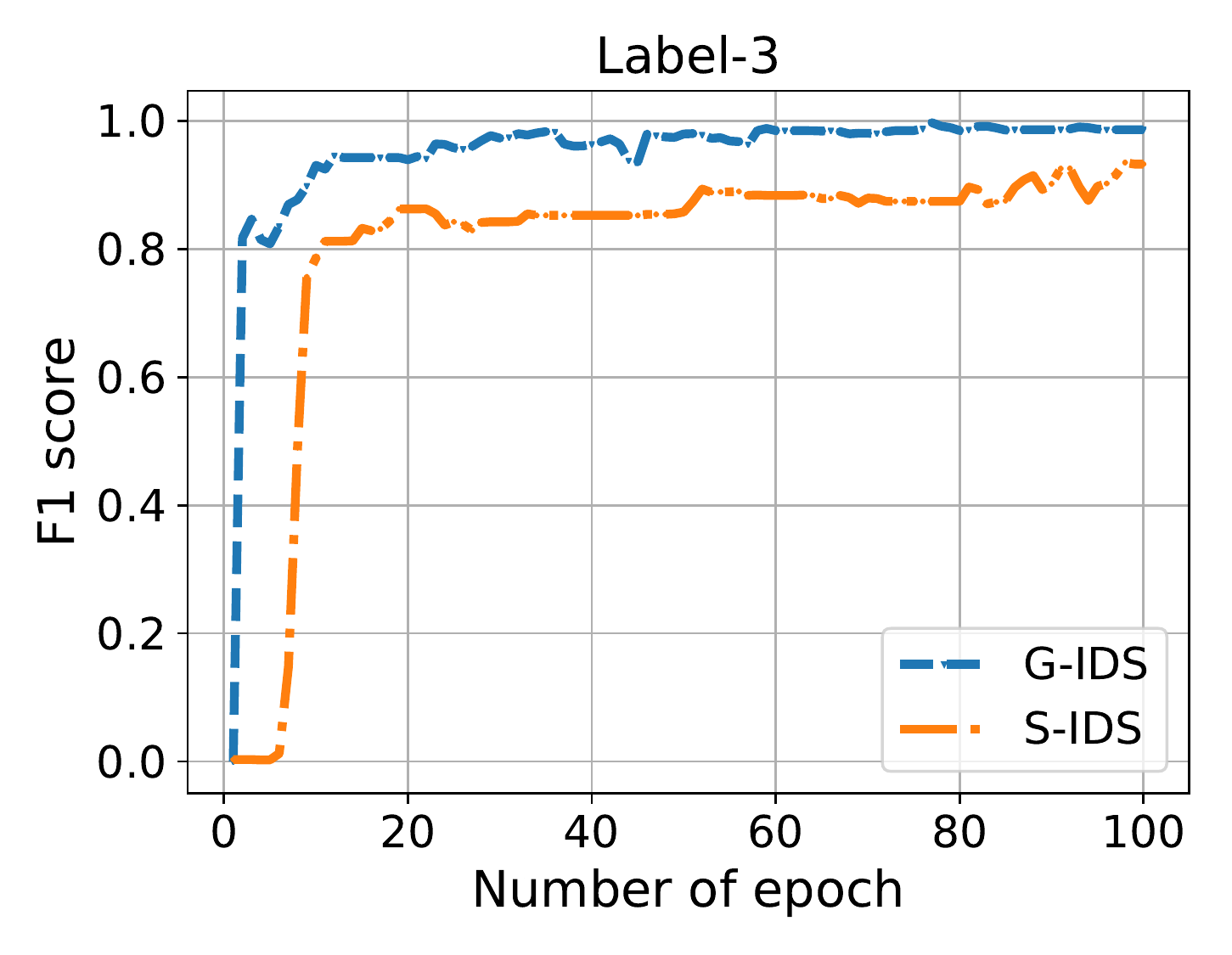}
        }
        \vspace{-5pt}
        \subfigure[]
        {
            \includegraphics[scale=0.45, keepaspectratio=true]{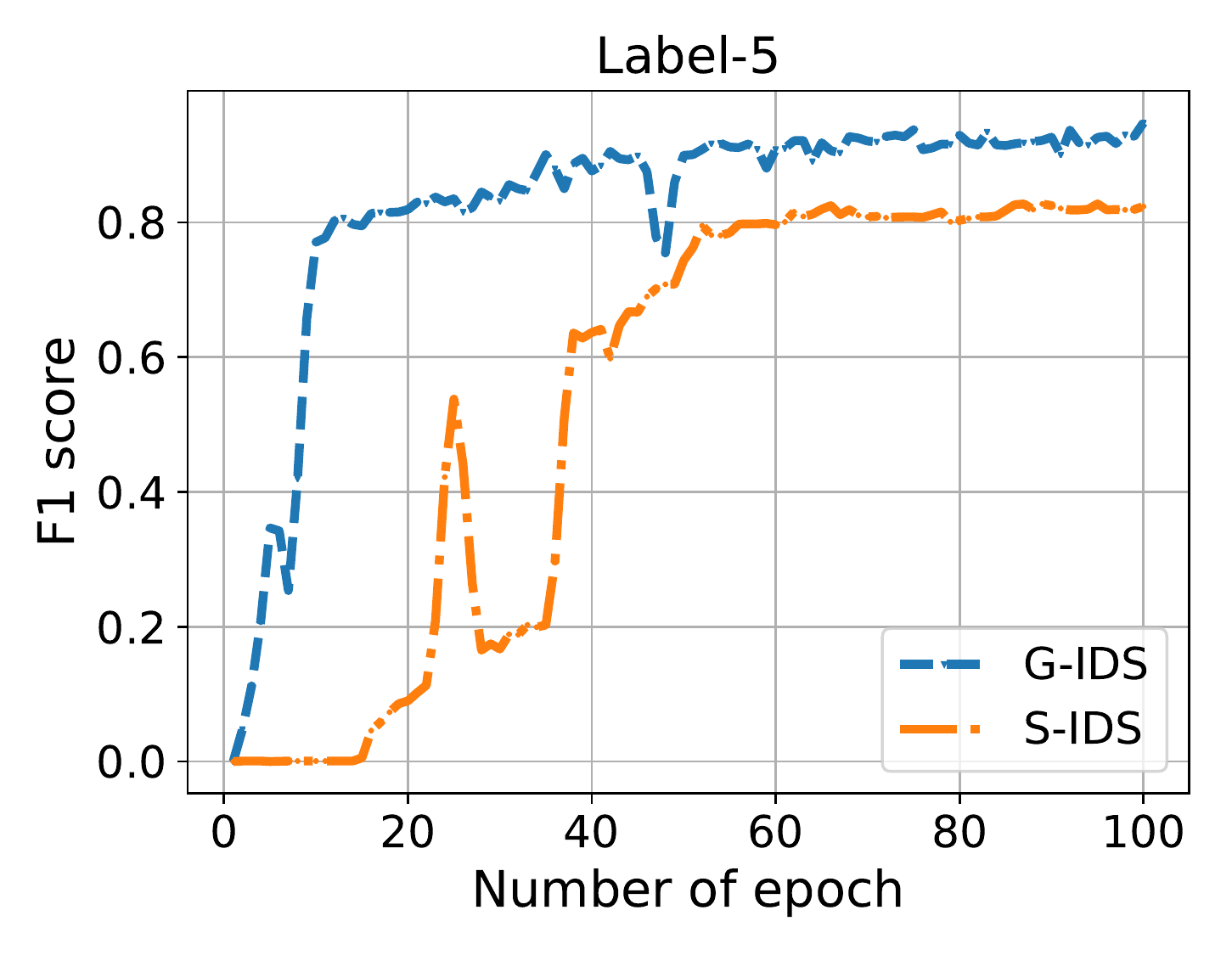}
         }
        \subfigure[]
        {
            \includegraphics[scale=0.45, keepaspectratio=true]{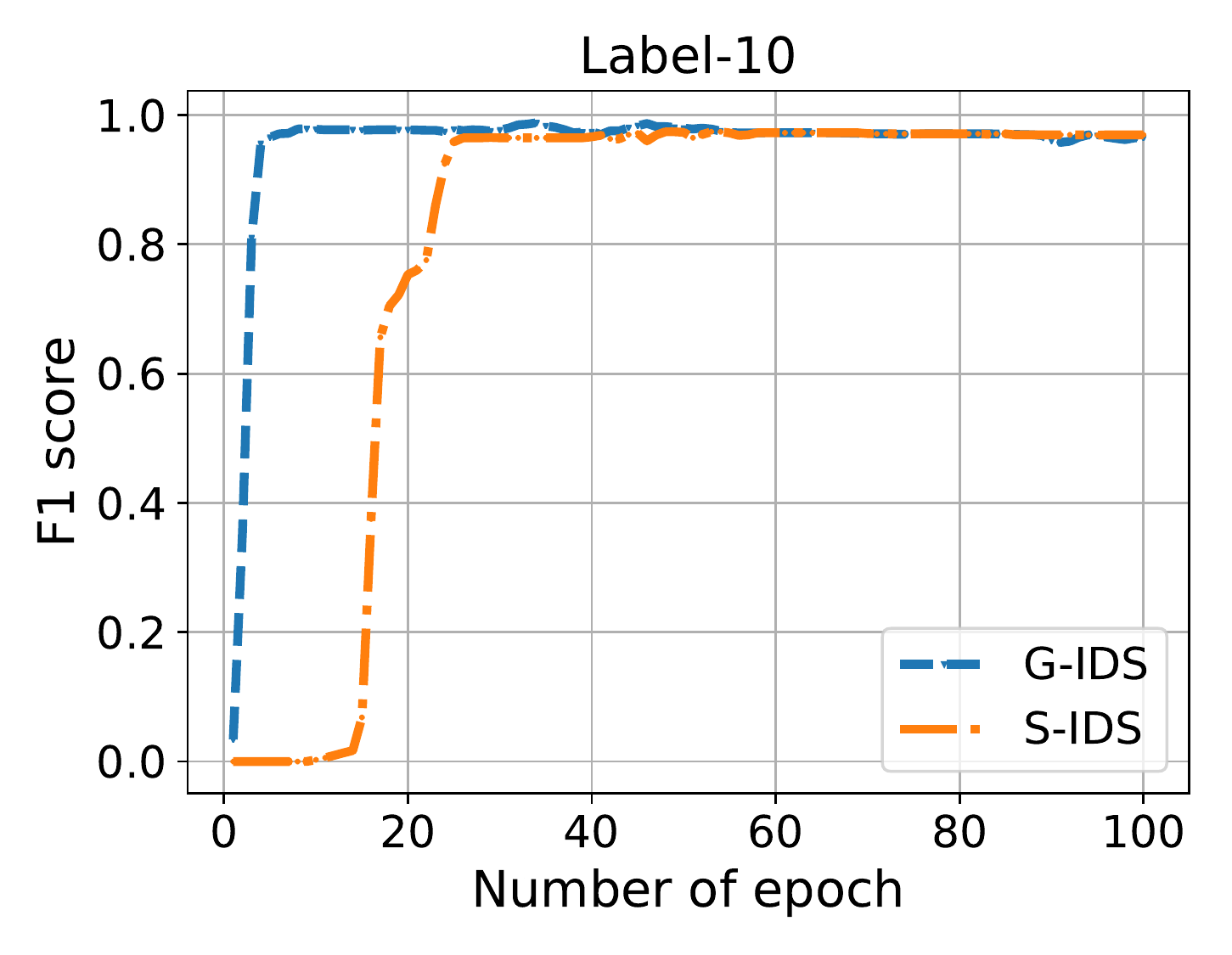}
         }
    \end{center}
    \caption{Stability analysis of S-IDS and G-IDS during training process for (a) label 1, (b) label 3, (c) label 5 and, (d) label 10.}
    \label{Fig:Eval_1}
    \vspace{-10pt}
\end{figure*}


\begin{table}[H]
\centering
\caption{Evaluation factors}
\label{tab:metric}
\vspace{-5pt}
\begin{tabular}{|l|l|l|}
\hline
\textbf{Indicator}   & \textbf{Actual Label} & \textbf{Prediction} \\ \hline
True Positive, TP  & Attack          & Attack              \\ \hline
False Positive , FP& Normal          & Attack              \\ \hline
True Negative, TN  & Normal          & Normal              \\ \hline
False Negative, FN & Attack          & Normal              \\ \hline
\end{tabular}
\vspace{-10pt}
\end{table}

Thus, from these four terms 
performance of an IDS can be easily calculated using the following equations \cite{el2016analysis}:
$$Precision,~~ P= \frac{TP}{(TP+FP)}$$
$$Recall,~~ R = \frac{TP}{(TP+FN)} $$
$$F1~score = \frac{2*P*R}{(P+R)} $$

As our dataset contains a lot of benign data, evaluating the system with a weighted average F1 score does not provide a clear picture. We consider the macro-average F1 score as it computes the metric independently for each label and then takes the average to treat all labels equally.
\section{Evaluation Results and Discussion} \label{sec:results}

The experiments are conducted on Dell Precision 7920 Tower workstation with Intel Xeon Silver 4110 CPU @3.0GHz, 32 GB memory, 4 GB NVIDIA Quadro P1000 GPU. The performance of G-IDS is evaluated in two perspectives, contrasting with the performance of a standalone IDS (S-IDS).

\subsection{Evaluation based on the Size of Training Dataset}
To investigate the performance of S-IDS and G-IDS, we split the entire training dataset into five parts. We take 20\%, 40\%, 60\%, 80\%, and 100\% of the training data to observe GAN's contribution with a growing dataset. 

\subsubsection{Label-wise Evaluation}
Firstly, we train S-IDS model without the assistance of GAN, which gives good results in the prediction for a few of the specific labels. However, S-IDS struggles to predict a few of the classes accurately as the number of training samples for those particular classes is not sufficient. As shown in Fig.~\ref{eval_5}, the F1 score of S-IDS for label 5 is approximately 0.70 while considering 20\% of data as the training data, and it increases gradually up-to 0.78 with the increasing training dataset.  However, training G-IDS with the GAN generated data along with the original data boosts the performance of G-IDS drastically. As shown in the figure, the F1 score of G-IDS reaches up to 0.90 for 20\% of data as training data and becomes 0.96 for 40\%. It is clear from the figure that G-IDS performs much better than S-IDS in attack detection. However, the fall of the F1 score for G-IDS at 60\% is not surprising as GAN's performance is highly random as it works by taking random noises as input. Similar improvements are also observed for label 6 and 9, shown in Fig.~\ref{eval_6} and  Fig.~\ref{eval_9}, respectively. Though there are improvements in the predictions for most of the labels, we only show these three labels as an example. 

\subsubsection{Overall Evaluation} 
In this part, we contrast the performance S-IDS and G-IDS on overall prediction for all the classes. As mentioned in the previous section, we use the macro F1 score as the metric, as it gives better insight into the model's performance on distinct labels. Fig.~\ref{fig:overl_all} represents the macro F1 score of S-IDS and G-IDS on the same test data while trained on different sizes of training data. We see that the macro F1 score of S-IDS improves from 0.82 to 0.87, as training data increases from 20\% to 100\%. On the other hand, G-IDS's performance improves by approximately 4\% as GAN-generated synthetic data are also considered in training. For the full training dataset, while S-IDS achieves a maximum macro F1 score of 0.87, G-IDS can take it at least up-to 0.91. 


Table~\ref{tab:ganidstable} summarizes the contrast of S-IDS and G-IDS performances when both are trained on 40\% of the original data. 
The first three columns represent the performance metrics of S-IDS for each class. 
If the F1 score of any class is less than the threshold, which we consider as 0.98 in this case, the controller asks the data synthesizer module to generate more synthetic data for that specific class to improve the performance of IDS. 

Initially, S-IDS has an F1 score of 0.85, 0.71, 0.55, and 0.33  for the labels of 2, 5, 6, and 9, respectively. These labels are considered as \textit{weak label} and are sent to the DSM to generate more training examples. 
GAN is trained on each of these labels and then generates 25\% new samples for each of these labels. Once verified by the controller, the newly generated synthetic samples are added to the existing hybrid dataset. The updated hybrid dataset is used to train the IDS again, and the whole process is continued. The second part of Table \ref{tab:ganidstable} shows the performance of G-IDS, which is trained on the hybrid dataset. From the results, we see that the detection rates of most of the classes are improved. To be more specific,  the F1 score of labels 2, 5, 6, and 9 increases up-to 0.93, 0.96, 0.68, and 0.41, respectively. 
Even though, by nature, GAN possesses the uncertainty of generating noisy unexpected data for one label that may overlap with the distribution of other labels, the controller module declines such kind of bad data and keeps the performance of G-IDS improving. 

\subsection{Evaluation of Improvement in the Training Stability}
As well as improving the final detection rate for different attacks, GAN also enhances the stability of the IDS's training process. Fig.~\ref{Fig:Eval_1} shows the performance of the model during training for four random targets. To present the performance of our proposed framework, we plot the F1 score for the test dataset for the increasing epoch. In the figure, we only show the first 100 epochs, which is sufficient to represent the contrast of the performance. From the figure, it is clear that S-IDS takes much more epochs to train the model and settle the F1 score to the final value. For S-IDS, most of the labels need approximately 50 to 100 epochs or even more to settle.

On the other hand, G-IDS does a pretty good job in the training stability. Almost all of the targets become stable within approximately 25 epochs. Thus the training process of G-IDS is much more stable than S-IDS. As the GAN-generated samples fill the gaps in the data distribution, it becomes easier for the IDS  to learn the distribution of the training data and get settled. Thus, in summary, G-IDS improves performance by balancing the imbalanced dataset as well as generating missing data. 
\section{Conclusion and Future Work} 
\label{sec:conclusion}
GAN is an influential tool in the deep learning area. IDS is also another fundamental agent for the CPS domain. In this paper, we propose a GAN supported IDS to unite them, which performs better than standalone IDS for an imbalance dataset or in any emerging field of cyber-physical systems where very few amounts of data are available for model training. We implement our model in a benchmark dataset, NSL KDD'99. Experimental analysis shows that our proposed G-IDS framework predicts with better accuracy than independent IDS, even after being trained with a small original dataset in the beginning. The centralized, computationally expensive and time consuming nature of G-IDS framework require further investigation. In future work, we aim to focus on creating a dynamic, efficient, and lightweight decentralized algorithm to implement it in the edge devices of the IoT realm. 
\bibliographystyle{unsrt}
\bibliography{g_ids}
%
\end{document}